# TIME IN THE 10,000-YEAR CLOCK

Danny Hillis[*], Rob Seaman[†], Steve Allen[‡], and Jon Giorgini[§]

The Long Now Foundation is building a mechanical clock that is designed to keep time for the next 10,000 years. The clock maintains its long-term accuracy by synchronizing to the Sun. The 10,000-Year Clock keeps track of five different types of time: Pendulum Time, Uncorrected Solar Time, Corrected Solar Time, Displayed Solar Time and Orrery Time. Pendulum Time is generated from the mechanical pendulum and adjusted according to the equation of time to produce Uncorrected Solar Time, which is in turn mechanically corrected by the Sun to create Corrected Solar Time. Displayed Solar Time advances each time the clock is wound, at which point it catches up with Corrected Solar Time. The clock uses Displayed Solar Time to compute various time indicators to be displayed, including the positions of the Sun, and Gregorian calendar date. Orrery Time is a better approximation of Dynamical Time, used to compute positions of the Moon, planets and stars and the phase of the Moon. This paper describes how the clock reckons time over the 10,000-year design lifetime, in particular how it reconciles the approximate Dynamical Time generated by its mechanical pendulum with the unpredictable rotation of the Earth.

## INTRODUCTION

The 10,000-Year Clock is being constructed (Figure 1 and Figure 2) by the Long Now Foundation.[**] It is an entirely mechanical clock made of long-lasting materials, such as titanium, ceramics, quartz, sapphire, and high-molybdenum stainless steel. The chamber for the pendulum-driven clock is currently being excavated inside a mountain near the Texas/New-Mexico border in the southwestern United States. Its mechanism will be installed in a 500-foot vertical shaft that has been cut into a mountain to house the clock. The clock is designed to maintain its long-term accuracy by synchronizing to the Sun, through heating of a sealed chamber of air by a shaft of light that shines into the mountain at solar noon. The air pressure change generated by this light drives a piston that adjusts the clock. Diurnal solar heating is also used to wind the weight that powers the pendulum and sun-tracking mechanism.[1, 2, 3]

The mechanical Clock is a digital computer with analog inputs and outputs. The principle inputs are the oscillations of the pendulum, sunlight falling on the solar synchronizer, and the pre-computed correction to solar time as realized in the equation-of-time cam. The outputs include an analog depiction of the sky in the orrery and a digital display of the Gregorian calendar date. These inputs and outputs are discussed below.

---

[*] Co-chair, Long Now Foundation, Fort Mason Center, Building A, San Francisco, CA 94123
[†] Senior Software Systems Engineer, National Optical Astronomy Observatory, 950 N Cherry Ave, Tucson, AZ 85719
[‡] Programmer-analyst, UC Observatories & Lick Observatory, Univ. of Calif., 1156 High Street, Santa Cruz, CA 95064
[§] Senior Analyst, Solar System Dynamics Group, NASA/Jet Propulsion Laboratory, California Institute of Technology, 4800 Oak Grove Drive, Pasadena, CA 91109
[**] http://www.longnow.org

**FIVE KINDS OF TIME GENERATED BY THE CLOCK**

Subtle differences in the definition of time can make a significant difference over 10,000 years. Universal Time will be $2 \times 10^5$ seconds behind Terrestrial Time (TT) in 10,000 years, because they tick different seconds. Barycentric Dynamical Time will be $5 \times 10^3$ seconds ahead of TT even though both tick *SI* seconds.[*] None of these time bases is more inherently correct than another. They are just different ways of labeling a sequence of instants in time. Over its 10,000 year life the display of the clock, averaged over the course of a year, is designed to remain within 300 seconds of Universal Time (UT).

How does the concept of UT affect the clock? As the word "average" suggests, Universal Time is a mean realization of solar time. The length of the solar day varies continually, seasonally and over the long term. In particular, there is a secular slowing due to lunar tides. By removing the seasonal variations, the implicit usage of UT[†] permits the regular pendulum-driven cadence of the clock to adapt to the long term changes.

The clock ultimately uses but does not explicitly display the solar system barycentric coordinate time of general relativity. This timescale is the independent variable in the equations of planetary motion that emerge from Einstein's space-time field equations and metric tensor. It is therefore a direct expression of our current understanding of the space-time relationship.[4, 5, 6] A defined relationship between coordinate time in the solar system barycentric frame and International Atomic Time (TAI) at a site on Earth (or Earth satellite) can be used to properly relate these timescales.

Current terminology of the International Astronomical Union (IAU) refers to the coordinate time of general relativity as Barycentric Dynamical Time (TDB) and, when properly related to a site on or near the surface of the Earth, Terrestrial Time (TT) can be derived. The distinction between these two dynamical timescales is generally periodic with an amplitude of 0.002 seconds in the course of a year, due to the Earth's elliptical orbit around the Sun. The self-correcting mechanisms of the clock described below mean the distinction is not significant operationally, and so "Terrestrial Time" will be used in the remainder of the paper, and the term "86400-second day" will be used refer to an 86400-second interval of Terrestrial Time.

With these underlying theoretical concepts in mind, there are five different time bases generated within the clock: Pendulum Time, Uncorrected Solar Time, Corrected Solar Time, Displayed Solar Time, and Orrery Time. Each will be described in detail.

---

[*] The *SI* second is defined based on specific transitions of a cesium atom. If this is measured on the rotating Earth, relativistic effects that depend in part on the earth's rotation will influence a cesium clock's rate relative to a clock in an inertial frame. Thus, even the measurement of the *SI* second that is used to define Terrestrial Time will be at least slightly dependent on the unpredictable rate of Earth's rotation, although the effect of variations in the Earth's rotation is not currently within the sensitivity of our measurements.

[†] The proposal to redefine Coordinated Universal Time (UTC) raises the issue of which concept will live longer, UTC or general-purpose "UT" with its meaning of mean solar time. The Clock will likely see many such cultural debates over its long lifetime. In the case of UTC no longer providing actual Universal Time, visitors' wristwatches will wander willy-nilly over the centuries in comparison to the Clock's solar-synchronized display.

**Pendulum Time**

The generative time base in the clock is called Pendulum Time. It is created by counting the 10-second cycles of the clock's gravity pendulum.[7] Pendulum Time advances every 30 cycles of the pendulum, once every five minutes. All other forms of time in the clock are derived from Pendulum Time and measured to this five-minute resolution. The mechanical pendulum is designed to have a long-term accuracy of better than 800 milliseconds per day, about one tick per year. The choice of a five-minute resolution is governed primarily by the requirement that the clock be robust through sunless periods lasting many decades. This also simplifies the design of the solar synchronizer and provides sufficient precision for the orrery display. The pendulum is driven by stored energy provided by a solar power mechanism or human winding. The clock is designed to operate unattended for as many as 10 millennia between human visits. The power system does not depend on sunny days, but only on diurnal temperature variations.

**Uncorrected Solar Time**

Uncorrected Solar Time is computed from Pendulum Time by adding a seasonally varying correction for the equation of time. This varies about half an hour or so over the course of a year. The analemma is a two-dimensional graph of the equation of time (Figure 3). The equation of time also varies from year to year, century to century. This variation is predictable within the uncertainty of the Earth's rotation rate, so it is pre-computed over the clock's 10,000-year operating interval and stored in the equation-of-time cam based on values is derived from an extended form of the JPL DE422 solar system solution The function encoded in the cam assumes the predicted slowing of the earth's rotational period at the rate of 1.8 milliseconds per day per century (Figure 4). The cam function also encodes the uncertainty of slowing of the Earth's rotational period, by a mechanism that will be described later.

The equation-of-time cam is driven by Corrected Solar Time, which is derived from Uncorrected Solar Time. Since this in turn is used to generate Corrected Solar Time, this sounds like a circular definition. It is, but because the corrections of Corrected Solar Time create a very small change in the equation of time correction, the "gain" around the loops is very small and the series converges.

Like Pendulum Time, Uncorrected Solar Time advances once every five minutes. Uncorrected Solar Time should be regarded as a purely internal time scale within the Clock as it retains the inevitable drift of the pendulum, corrected in the next step.

**Corrected Solar Time**

Corrected Solar Time is a realization of local apparent solar time at the site of the Clock. It is intended to stay synchronized over the long-term with the rotation of the Earth. The solar synchronizer adds the required correction automatically whenever the Uncorrected Solar Time deviates from apparent solar time by more than five minutes.

Corrected Solar Time, having ticks straddling solar noon, is computed from Uncorrected Solar Time by adding or subtracting a correction tick if the sun synchronizer detects solar noon while the Uncorrected Solar Time is not within the noon interval. This solar synchronization corrects for both the inaccuracy of the pendulum and the unpredictable component of the Earth's rotation.

Thus, corrected solar time is the start time, plus the total number of pendulum ticks, plus the equation of time correction, plus the sum of the signed correction ticks. The correction is positive if the Sun is detected before the just-before-noon tick, and negative if it is detected after the just-after-noon tick. No correction is made if the Sun is detected in between these two ticks. On any given day a maximum of a single correction tick will be added or subtracted. Thus, if the Clock is ever in a state many minutes or hours from the correct local apparent solar time, it may take multiple sunny days to correct itself.

Because the correction may be negative, Corrected Solar Time as generated is not monotonic; it can go backwards, repeating a short interval of time. To prevent the mechanisms that are driven by Corrected Solar Time from moving backward, a mechanism for storing "borrowed-time" is used to stop the advancing shaft until the time "catches up," creating a monotonic version of Corrected Solar Time that pauses rather than backing up. Since only a single correct tick is added per day, the mechanism will not cause the clock to pause for more than five minutes.

All displays on the clock are designed to maintain accuracy to within a five-minute tick over the entire 10,000-year lifetime of the clock, as long as the clock detects solar synchronization at least once a year. It is possible that an unusual event such as a volcanic eruption could prevent the clock from detecting noon for more than a year. In this case, the clock may temporarily drift away from the correct time, but it will eventually resynchronize when clear skies reemerge as long as it has not drifted more than 12 hours. This allows the clock to successfully recover after more than a century of overcast skies. In that case it will require at least as many sunny days to recover the correct solar time as the number of 5-minute ticks the clock has drifted, 12 days per hour of drift. If the skies are overcast for several centuries, the Clock's calendar display could gain or lose days in its calendar count, but the correct solar time will be restored when the Sun returns.

**Displayed Solar Time**

Displayed Solar Time is generated from Corrected Solar Time each time the clock is wound by a visitor. The solar power mechanism is not used to advance this part of the mechanism. All of the clock's displays are derived from Displayed Solar Time, or Orrery Time, which is derived from it. Displayed Solar Time pauses between windings. It moves forward each time the clock is wound until it matches the monotonic version of Corrected Solar Time. The displays move forward as Displayed Solar Time moves forward, so the displays move only while the clock is being wound. Depending on how long it has been since the clock has been wound, winding may take a few minutes or many hours. The winding mechanism will stop the winding when Displayed Solar Time matches Corrected Solar Time.

Displayed Solar Time drives the portion of the dial that shows the apparent position of the Sun in the local sky. The displays indicate Displayed Solar Time in increments that straddle solar noon, so the just-before-noon tick, and the just-after-noon tick are displayed as 2.5 minutes before and after solar noon, respectively. Displayed Solar Time also drives a digital calendar display, which displays the exact date according to the Gregorian calendar system.[8]

**Orrery Time**

Orrery Time is used to compute the astronomical display of the position and phase of the Moon, the tropical year, the sidereal day, orbits of the visible planets, and the precession of the Earth's axis. Orrery Time is intended to approximate the solar system barycentric coordinate time. In that sense it is similar in purpose to Barycentric Dynamical Time (TDB), or the older Ephemeris Time (ET). Orrery Time differs from Corrected Solar Time because the rate of the Earth's rotation is slowing. Historical trends suggest that the average day is currently slowing by about 1.8 milliseconds per day per century. If this trend continues creating a deviation that grows quadratically with time, a Terrestrial Time clock, measuring an 86,400-second day, would differ from Corrected Solar Time by about 3.8 days after 10,000 years. This would not create noticeable error on the displays of the planetary positions, but it would be apparent on the display of the position and phase of the Moon.

Orrery Time is calculated from Displayed Solar Time by subtracting out a correction for the slowing of the Earth's rotation. Orrery Time is automatically corrected for the expected slowing of the Earth's rotation by a cam with a quadratic correction included in the function encoded on the cam. An additional mechanism is also provided for future adjustment to Orrery Time to match the observed slowing, as described below. No attempt is made to subtract the equation of time to produce a daily approximation of Terrestrial Time, since this error is too small to be seen on the display, and the length of the day will average out to the mean solar day over the course of each tropical year.

## DISPLAYS

### Days

Three types of days are computed for driving the displays: solar days, orrery days and sidereal days. The clock also makes use of the 86400-second day in the design of its cams and gear ratios, but does not represent it explicitly in the displays.

Solar days are demarcated by the motion of the Sun display, which indicates the approximate position of the Sun in the sky (Figure 5 and Figure 6). The Sun travels around a circle divided by a horizon line that adjusts throughout the tropical year so the Sun will show the time of sunrise and sunset. Solar days are also counted on the calendar display, which indicates day of week, calendar month, and day of month. The Sun display and calendar date are both driven directly from Displayed Solar Time.

Orrery Days are the clock's long-term approximation of 86400-second days (TT), used to drive the motions of the planets in the orreries. They are derived from Orrery Time. They differ from the 86400-second days in that they vary in length with the season, but they closely approximate 86400-second days when averaged over the course of a tropical year.

Sidereal days are used to drive the horizon-line indicators that move across the star field display. The horizon-line indicators are similar in form to the "rete" of an astrolabe, although on a classical astrolabe the stars move across the background representing the horizon.[9] In the clock, the horizon indicators move across the background of the stars. The clock sidereal day is produced by adding the sidereal year rate to the motion of the Sun to produce the sidereal days. Since the sidereal year rate is derived from Orrery Time and the motion of the Sun is derived from Displayed Solar Time, the rate of the horizon-line indicator is derived from a combination of both of the time bases.

### Years

Four different versions of the year are generated by the clock: the tropical year, the sidereal year, the calendar year and the cam year.

The tropical year, that is the year corresponding to the seasons, is used to compute the horizon dial that adjusts the length of the displayed day for the time of year. The number of 86400-second days in a tropical year is 365.2421896698-0.00000615359 $T$, where $T$ is the number of Julian centuries past the year 02000[*].[10] The average length of the tropical year, over the next 10,000 years, will be 365.241882 86400-second days (TT). The tropical period is used to drive the cam that moves the horizon dial. It is calculated from the Orrery Time.

---

[*] The Long Now Foundation uses five-digit dates, the extra zero is to solve the deca-millennium bug which will come into effect in about 8,000 years.

The sidereal year is the time it takes the Earth to move once around the Sun with respect to an inertial coordinate system (formerly the "fixed" stars, now a similar, agreed-upon coordinate system defined by the compact radio sources of the International Celestial Reference Frame (ICRF). The sidereal year is used to display the position of the Earth in the orrery, and its calculation is discussed in the section describing the calculations of the orbits of the planets. It is also used to compute the sidereal day to drive the horizon line indicators in the star display. The sidereal year is derived from Orrery Time.

The calendar year is computed according to the Gregorian calendar system, including the leap-year exceptions for centuries and millennia. There is still some question about whether to incorporate a possible future reform of the Gregorian calendar[11], in which the millennial exception skips every fourth millennium. This would keep the calendar in closer synchronization with the tropical year. The current thought is not to do so, on the grounds that the Gregorian calendar is displayed primarily as a cultural artifact of our time. The calendar year is displayed in a five-digit format.

The equation of time and solar elevation cams are rotated once a "cam year," which is defined to be 365.2222 mean solar days. Since the two-dimensional functions encoded in the equation-of-time cam are actually derived from a one-dimensional time varying function, there is flexibility in the exact ratio of days to cam years. This interval was chosen because the ratio is easily computed by gears and makes a beautifully shaped cam. It is calculated from Corrected Solar Time by gear trains with a ratio of $2958/81 = 365.2222$.

**Moon**

The face of the dial displays the phase of the Moon and the position of the Moon in the sky. The Moon position display is an aperture that moves over the background of a 16 lunar phase display that moves with the Sun display. The aperture normally moves with this background at the same rate as the Sun, but it occasionally moves backwards against it, in discrete steps of 1/16 full rotation, changing its phase with respect to the Sun display. The displayed phase of the Moon is determined by the angle between the Sun display and the Moon display. The Sun display is driven from Displayed Solar Time and the Moon aperture is driven from the sum of the Sun rate and the backwards steps. These backward steps are derived from Orrery Time, so the Moon display is computed by a combination of Orrery Time and Displayed Solar Time.

The phase of the Moon is a very sensitive indicator of the long-term ratio of the month and day lengths, which depends not only on long-term changes in the rotation of the Earth, but also long-term changes in the orbit of the Moon. These are caused primarily by tidal dragging of the Earth's oceans, landmasses and atmosphere. The mean sidereal month over the next 100 centuries is 27.3216719 86400-second days. As explained below, this mean value is derived from an extended form of the JPL DE422 solar system solution. The number of 86400-second days in the mean synodic month (average time from new moon to new moon) is derived by subtracting the mean sidereal period of the Earth's orbit from the mean sidereal period of the Moon.

**Stars**

The face of the dial shows the position of the stars in the sky, which rotate once per sidereal day. The sidereal day rate is equal to the solar day rate plus the sidereal year rate. In the clock this is generated by the motion of the horizon indicator over the star field, which in turn rotates at the precession rate of the Earth's axis. Both of these are generated from Orrery Time. The horizon should rotate in 0.9972696693 solar days, which is about 23 hours 56 minutes. This is the mean period of the sidereal day over the next 10,000 years, given the predicted slowing.

The stars display also takes the precession of the equinoxes into account. This rotation happens about once every 26,000 years, so there will be less than half a rotation in the lifetime of the clock. In 10,000 years, the bright star Vega in the constellation Lyra will be the pole star. Vega is currently known as one corner of the so-called "Summer Triangle." At the end of the planned 10,000-year design life of the clock, it will be a circumpolar star visible in all seasons of the year.

The clock also generates the precessional rate from Orrery Time. Because the star display is generated by horizon line indicators moving across the star field, and the horizon line indicators are driven in part by Displayed Solar Time, the star display is computed by a combination of Orrery Time and Displayed Solar Time.

**Planets**

The clock will display the visible planets in two orreries: a heliocentric Copernican orrery (Figure 7 and Figure 8), and a geocentric Ptolemaic orrery. For this purpose the sidereal orbital periods are used, modeling the mean motion of the planets, rather than the true elliptical orbits. The Ptolemaic orrery is driven from these same periods, with a mechanical system that computes the appropriate coordinate transformation. The planetary orbit rates are all derived from Orrery Time.

**Table 1. Time Periods displayed by the clock. The "Gear Ratio" column shows the ratio of the gear train connecting the display to the indicated time base, approximating the number of the Predicted Solar Days.**

|  | *86,400-Second Days (TT)* | *Predicted Solar Days* | *Gear Ratio* | *Time Base* |
|---|---|---|---|---|
| Solar Day | 1.000001042 | 1 | 1 | Solar |
| Sidereal Month | 27.32167193 | 27.32164347 | 27.3216438 | Orrery |
| Synodic Month | 29.53060095 | 29.53057019 | 29.53057085 | Orrery |
| Sidereal Day | 0.997270708 | 0.997269669 | 0.997269669 | Orrery |
| Precession | 9412982.24 | 9412972.435 | 9412882.619 | Orrery |
| Tropical Year | 365.242189 | 365.2418085 | 365.2415166 | Orrery |
| Sidereal Year | 365.2563681 | 365.2559877 | 365.2564103 | Orrery |
| Cam Year | 365.2226027 | 365.2222222 | 365.2222222 | Solar |
| Mercury | 87.96925644 | 87.96916481 | 87.96914701 | Orrery |
| Venus | 224.7007992 | 224.7005652 | 224.7037994 | Orrery |
| Earth | 365.2563681 | 365.2559877 | 365.2564103 | Orrery |
| Mars | 686.9798408 | 686.9791252 | 686.978544 | Orrery |
| Jupiter | 4332.599090 | 4332.594577 | 4332.798497 | Orrery |
| Saturn | 10759.08080 | 10759.06959 | 10755.49679 | Orrery |

Table 1 shows the sidereal periods of the planets in 86,400-second days, and in predicted mean solar days of 86,400.09 seconds, averaged over the next 10,000 years, based on the projected slowing of the solar day. The sidereal planetary periods, as well as those of the sidereal month, are derived from an extended form of the JPL Development Ephemeris 422 (DE422) solar system solution and the Horizons software and algorithms.* DE422 is a slightly updated version of DE421.[12] This planetary ephemeris is an integration of 2nd-order parameterized post-Newtonian (PPN) n-body equations of motion consistent with general relativity, fit to several centuries of accumulated ground-based measurements and spacecraft tracking data. When these fundamental equations are numerically integrated, many periodicities and secular trends in planetary motion emerge as a result of mutual perturbations such as resonances. Therefore, to properly determine a mean sidereal period, the ephemeris for each object was sampled at intervals between 1 and 50 days (depending on object) over the 10,000-year interval of interest (02011-12011). A single precessing ellipse was then estimated for each planet's dataset in a least-squares sense, iteratively converging on a parameter set that is a least-squares best fit of an ellipse to the computed dataset. The resulting mean sidereal periods above are thus the statistically optimal fit for the full 10,000-year timespan of the clock, relative to the DE422 solution. If a different timespan was considered, slightly different values would result.

**DEALING WITH THE UNPREDICTABLE VARIATIONS OF EARTH'S ROTATION**

As previously discussed, the Earth's rotation is currently slowing at a rate of about 1.8 milliseconds per day per century.[13] Of course, the trend may not continue, especially if the climate changes. The variation is caused by a variety of effects including tidal drags, shifts in the Earth's crust, changes in ocean levels, and even weather (Figure 9. For example an ice age would put more mass near the poles, making the day shorter. Melting icecaps would make it longer.[14] This creates an uncertainty in the average length of day of about 10 parts per million, an uncertainty of plus or minus 37 solar days over the design lifetime of the clock.†

Since the difference between Orrery Time and Displayed Solar Time is only clearly visible in the Moon display, determining the correct value is not crucial for overall functioning of the clock. For this reason, provision is provided for user adjustment, allowing future observers to tune the Moon display to match their observations of the Moon. The clock will record the history of these observation-based adjustments, making the 10,000-Year Clock a scientific instrument for recording the long-term slowing of the Earth's rotation.

Human observation is a satisfactory method of converting Displayed Solar Time to Orrery Time for display, but it cannot be depended on for the operation of the timekeeping mechanisms of the clock. The displays only matter when they are visited by humans, but the timekeeper is designed to continue to keep track of the correct time without human intervention. There is only one mechanism in the timekeeping of the clock that is sensitive to the unpredictable slowing of the Earth's rotation: Corrected Solar Time is used to drive a light steering prism that guides the noontime light into the mountain. This steering prism is required to steer the light down a 500-foot shaft to the light sensor, adjusting the Sun's varying noontime elevation during the course of the tropical year.

The solar elevation is calculated from a cam that already encodes the predicted slowing of the Earth. Our uncertainty in this prediction translates to a growing uncertainty in the elevation of the noontime Sun. The Sun elevation cam encodes this uncertainty by seeking for the Sun in different positions of the uncertainty window on successive days, scanning the entire uncertainty window back and forth so that it will always be within the acceptable accuracy for at least one day during each scan. This works because the clock does not require a synchronization event every day. The scans are performed twelve times every tropical year. Since the uncertainty scanning grows over time, the window of uncertainty over which the scans are performed goes accordingly. Thus, the scanning deviates very little from the predicted value in the early centuries, but the perturbations become more noticeable as time progresses.

---

* Giorgini, J.D., NASA/JPL Horizons On-Line Ephemeris System, 2011 (http://ssd.jpl.nasa.gov/?horizons).
† Busch, M.W., "Climate Change and the Clock." (http://blog.longnow.org/category/clock-of-the-long-now/page/2/.)

Since the equation of time is also sensitive to deviations in the phase of the tropical year, the function encoded in the equation-of-time cam can also encode the same time scanning as the Sun elevation cam. Thus, when the Sun is detected, Uncorrected Solar Time will be adjusted with the appropriate equation of time value to create corrected solar time.

**CONCLUSION**

Human societies have always organized their activities around the rising and setting of the Sun. Civilization required agriculture. Agriculture required sunlight. Much of human culture is organized around a diurnal or annual cadence. Systems whose primary duty cycles are much shorter or much longer than a day will still have some superposed diurnal signature resulting from human-mediated interactions, such as maintenance and administrative activities. In this case, both the solar-powered and regulated mechanism of the Clock, as well as its concept of human-mediated operations to update the orrery and calendar displays, will exhibit strong diurnal peaks of activity. The Sun matters to humans, even to their devices in the depths of space and on the surfaces of other planets. It has done so throughout history and the Clock makes the statement that it will continue to matter thousands of years into the future.

Yet the Clock is more than a sundial. It also represents the apparent positions of other astronomical objects: the Moon, the planets and the stars. Unlike a sundial, the representation of apparent solar time becomes an engineering choice, not the result of directly measuring the hour angle or azimuth of the Sun from its shadow. To implement a living simulation of the solar system in a shaft drilled 500 feet into a mountain requires precisely the same dynamic conversation between a carefully tuned physical oscillator and the changing syncopations of the natural world In the case of the 10,000-Year Clock, the former is its pendulum and the latter the orrery. In the case of civil timekeeping, it is an ensemble of climate controlled atomic chronometers versus the rotation of the Earth as represented by Universal Time.

As the centuries pass the occasional resynchronization of the Clock on some sunny day will correct for the slight residual drift of the high-precision pendulum. But it will also accommodate the quirky residuals in the Earth's rotation remaining after compensating for the pre-computed equation of time. These residuals are precisely the result of variations in the mean solar length-of-day as shown in Figure 9 Like all conscientious timekeepers the clock eventually just reconciles the steady dynamical swing of its pendulum with the wavering rotation of the Earth.

**ACKNOWLEDGMENTS**

The authors thank William Folkner (JPL) for extending the DE422 planetary ephemeris integration to cover the clock's entire interval of operation and Robert Jacobson (JPL), whose natural satellite software was adapted to compute the mean sidereal periods of the planets and Moon over this timespan, and Michael Busch for his help in making the astronomical calculations required for the design of the clock.

**APPENDIX: FIGURES**

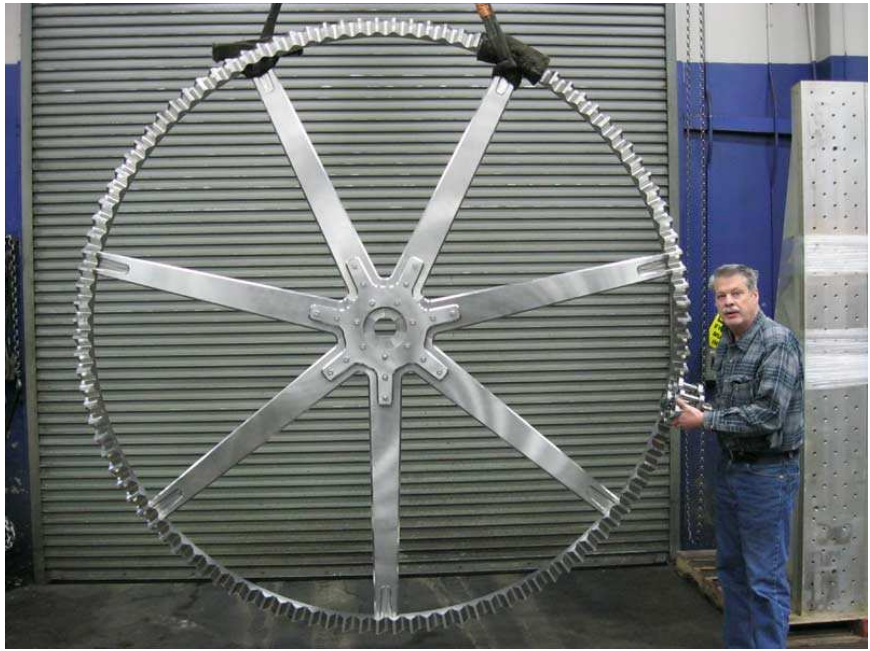

**Figure 1. A drive gear to be used in the 10,000-year clock.**

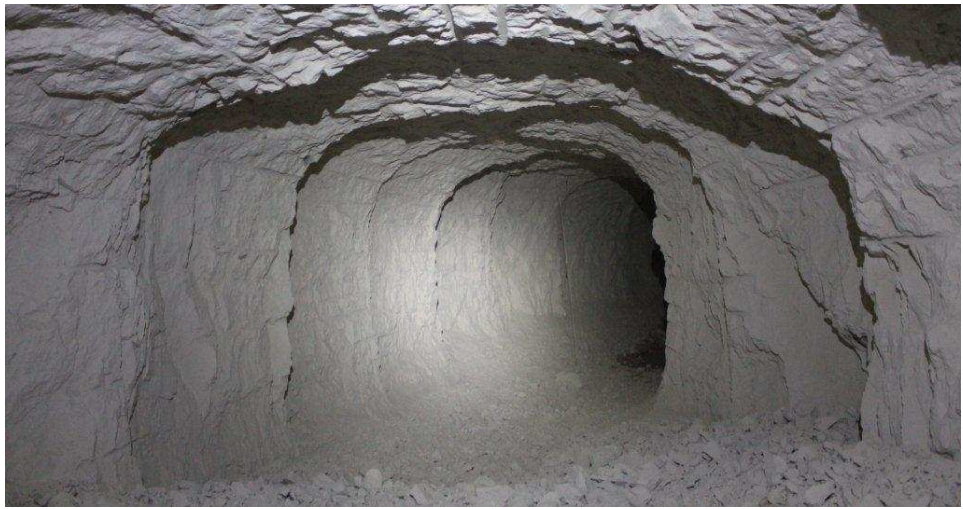

**Figure 2. Tunnel that is currently being excavated to reach the underground site of the clock**

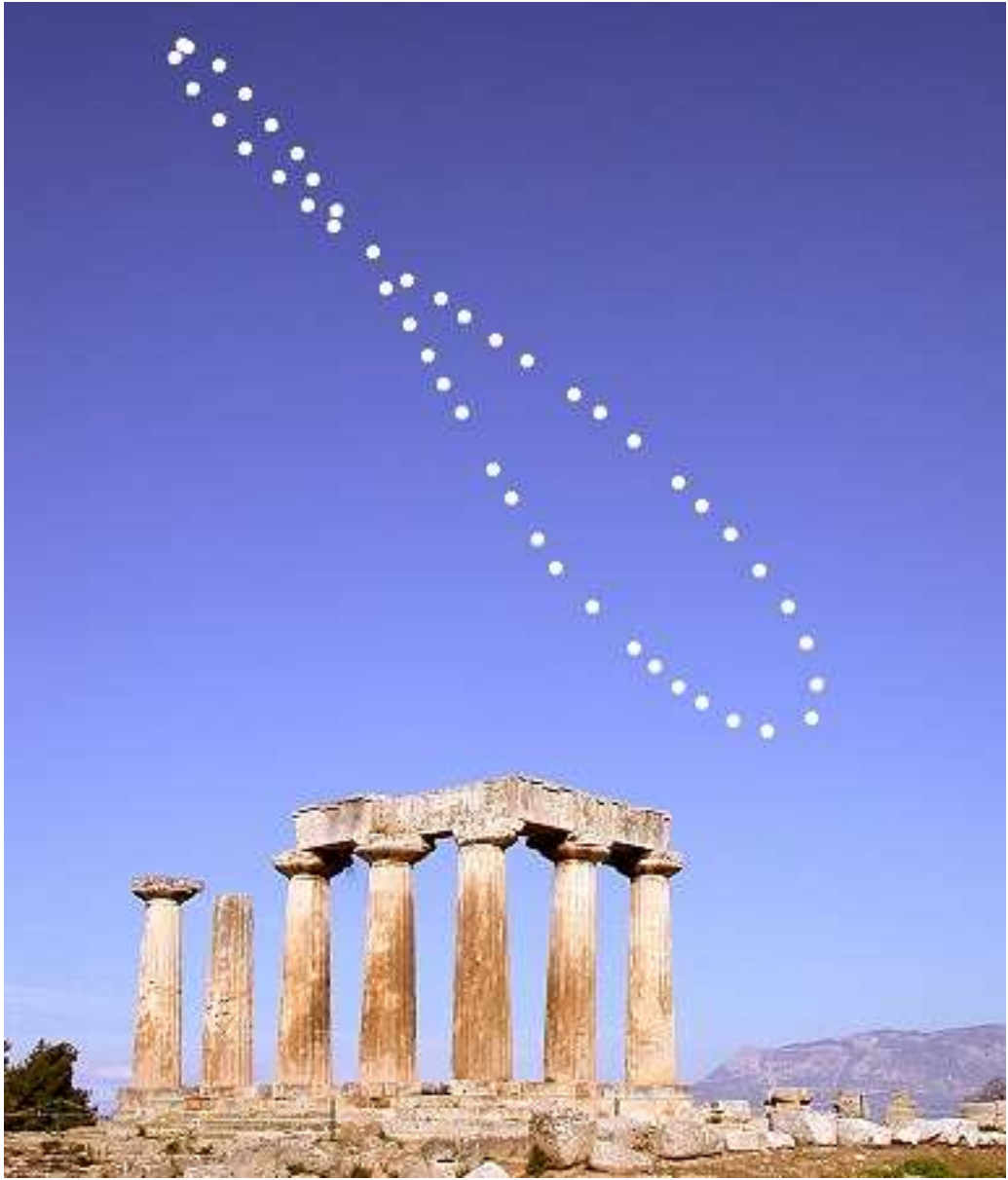

**Figure 3. Analemma with the Temple of Apollo, photographed by Anthony Ayiomamitis by making multiple exposures at 09:00:00 UT+2, on multiple days between January 7 and December 20, 2003.**[*]

---

[*] http://www.perseus.gr/, *A Vibrant Universe in Vivid Colour: Astrophotography by Anthony Ayiomamitis* (used with permission).

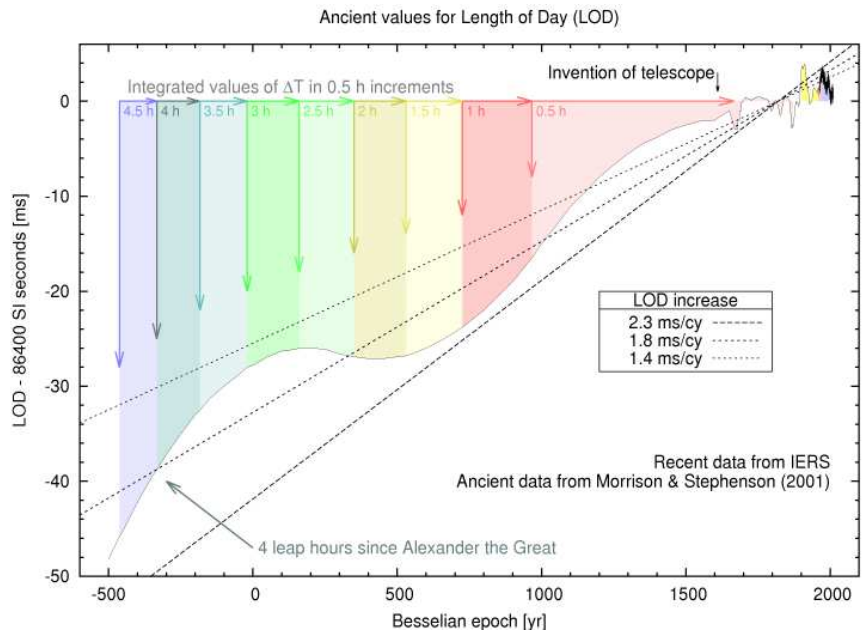

**Figure 4. Historical variation of the length of the day.**

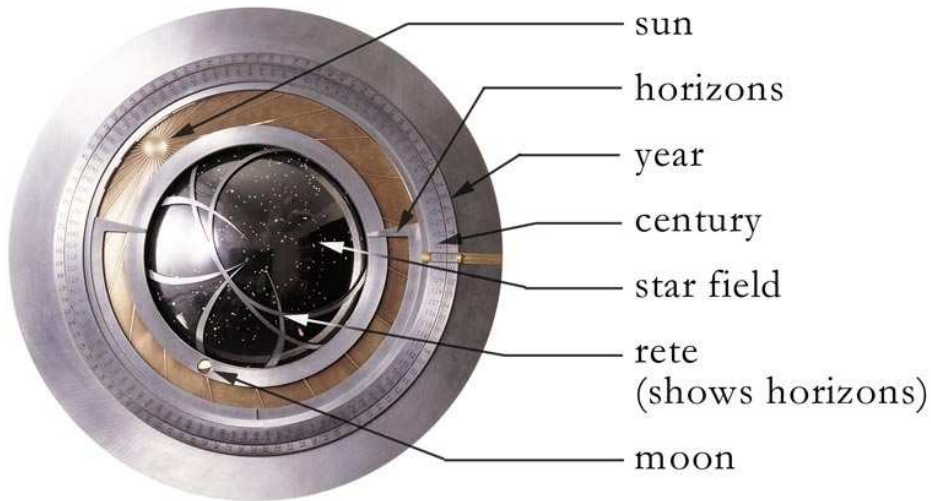

**Figure 5. The face of the prototype 10,000-Year Clock, showing the positions of the Sun, Moon, and stars. The actual 10,000-Year Clock will have a similar design, except that the calendar year and century will be on separate displays.**

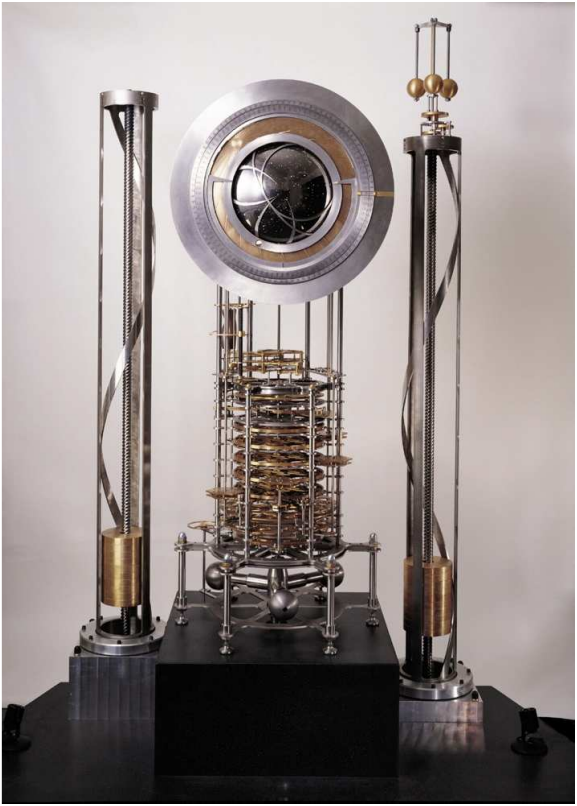
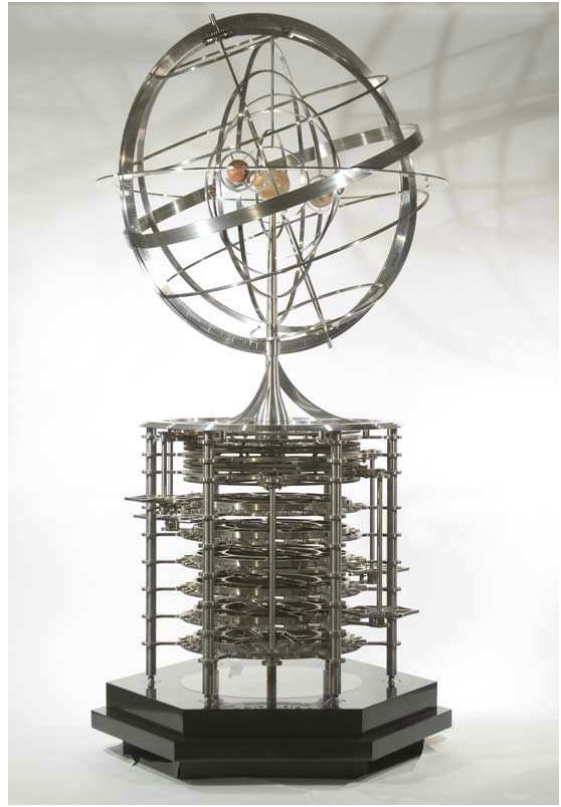

**Figure 6. Prototype 10,000-Year Clock, currently on display in the London Science Museum.**

**Figure 7. Prototype of the 10,000-Year Clock's Copernican orrery, currently on display at the Long Now Foundation's museum in San Francisco, CA**

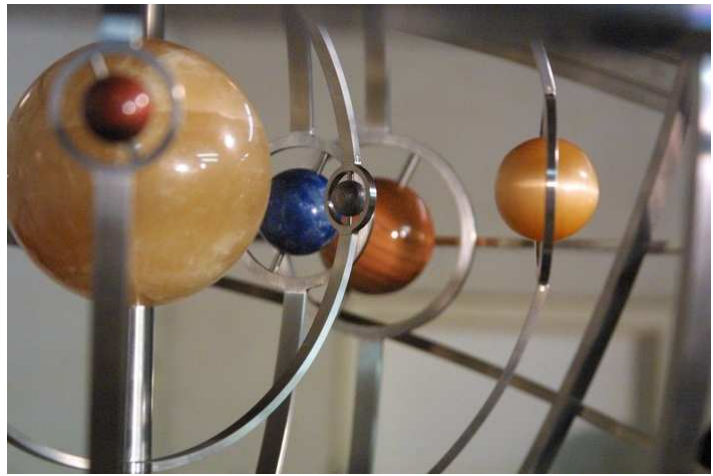

**Figure 8. Close up of the prototype of the 10,000-Year Clock's Copernican orrery, showing the stone spheres that represent the visible planets.**

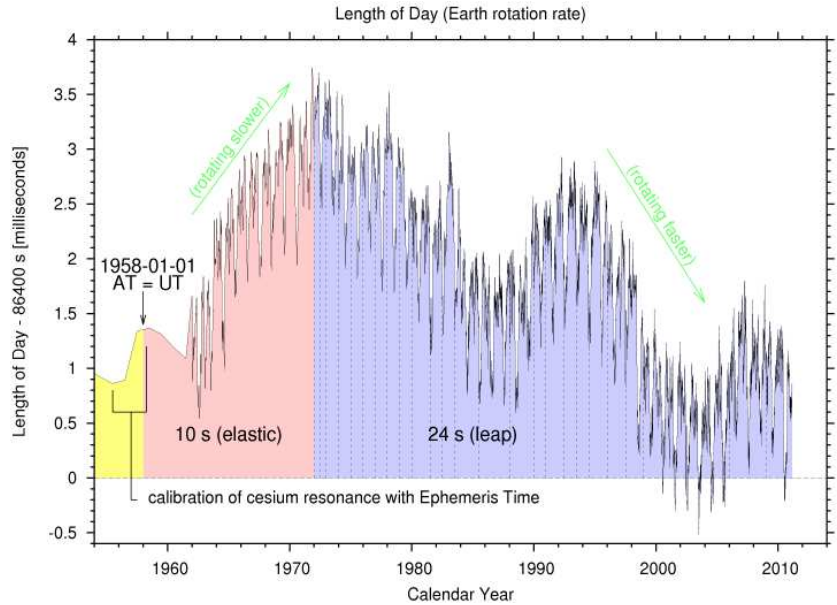

**Figure 9. Recent variations in Length of Day from IERS data. The number of seconds that must be added into the atomic time scale (AT) to track Universal Time (UT) is proportional to the area under the curve.**